\documentclass[10pt,a4paper]{article}
\usepackage[utf8]{inputenc}
\usepackage[T1]{fontenc}
\usepackage{amsmath}
\usepackage{amsfonts}
\usepackage{amssymb}
\usepackage{graphicx}
\author{Hans Wilschut\footnote{University of Groningen\protect \\
		Email: hwwilschut@gmail.com
		\protect \\ home address:  Sankt Augustin, Germany}}
\title{Frequency shifts of head echos \\
	 in meteoroid trail formation}

\begin{document}\maketitle
\begin{abstract}
	The approximate  frequency shift of  a radar echo  from  a meteoroid  is derived. The origin of head echos are discussed by considering schematic models.
\end{abstract}
\section{Introduction}	
Many aspects concerning radio echos from meteoroid trails can be understood in terms of the line-oscillator model. The standard work for this is McKinley's book \cite{McK},  chapter 8 and 9. However, McKinley does not show how frequency shifts near the optimal reflection point can be calculated. In this paper the formalism of \cite{McK} is extended to include frequency shifts. In doing so a close relation appears between the line-oscillator and what is referred to as the "moving ball" model. In the latter model frequency shifts occur because the radio echo will be Doppler shifted. 

Inspecting measured meteoroid radio echo's  such as shown in figures~\ref{ex1} and \ref{ex2} one can observe two distinct parts.  The first part is a signal with a frequency  deviating from the transmitter frequency  and the second part,which extends over longer time, is centered around the transmitter frequency. This first part is often referred to as the "head echo".  
Many authors interpret this signal in terms of Doppler shifts. 
The  larger second part is well understood as a reflection from the trail left behind after a meteoroid passed. The free thermalized electrons in the trail can cause strong reflections of the radio signal when their individual  amplitudes add coherently.  
The head echo is produced during trail formation. In the line-oscillator description the scattering electrons, created as  the trail is formed, will have relative phases such  that it appears as a frequency  shift away from the emitter frequency.  This is not the Doppler shift due to reflections on a co-moving plasma. The latter is assumed  in the moving ball description. Some recent work on Doppler shifts in head echos can be found in this journal.  See for example \cite{Kauf,Verb, German}.

One aim of this paper is to derive an approximate value of the frequency shift using the line-oscillator model of \cite{McK}. The notation of that work will be followed unless otherwise indicated. For the convenience of the reader some of the formalism in ref.~\cite{McK} will be repeated here. The derivation of the frequency shift in this model will be given in the next section. The third section considers the case of Doppler shifts, showing the close connection with the line-oscillator model. In section 4  
example calculations are made for a qualitative comparison with observations.  Section 5 gives a suggestion why spectra like the one shown in fig.~\ref{ex1} with only a half head echo are seen more frequently than the one in fig.~\ref{ex2} with a complete head echo.   The final section contains some concluding remarks.

\section{Derivation of the frequency shift}
\begin{figure}
	\includegraphics[width=0.8\linewidth]{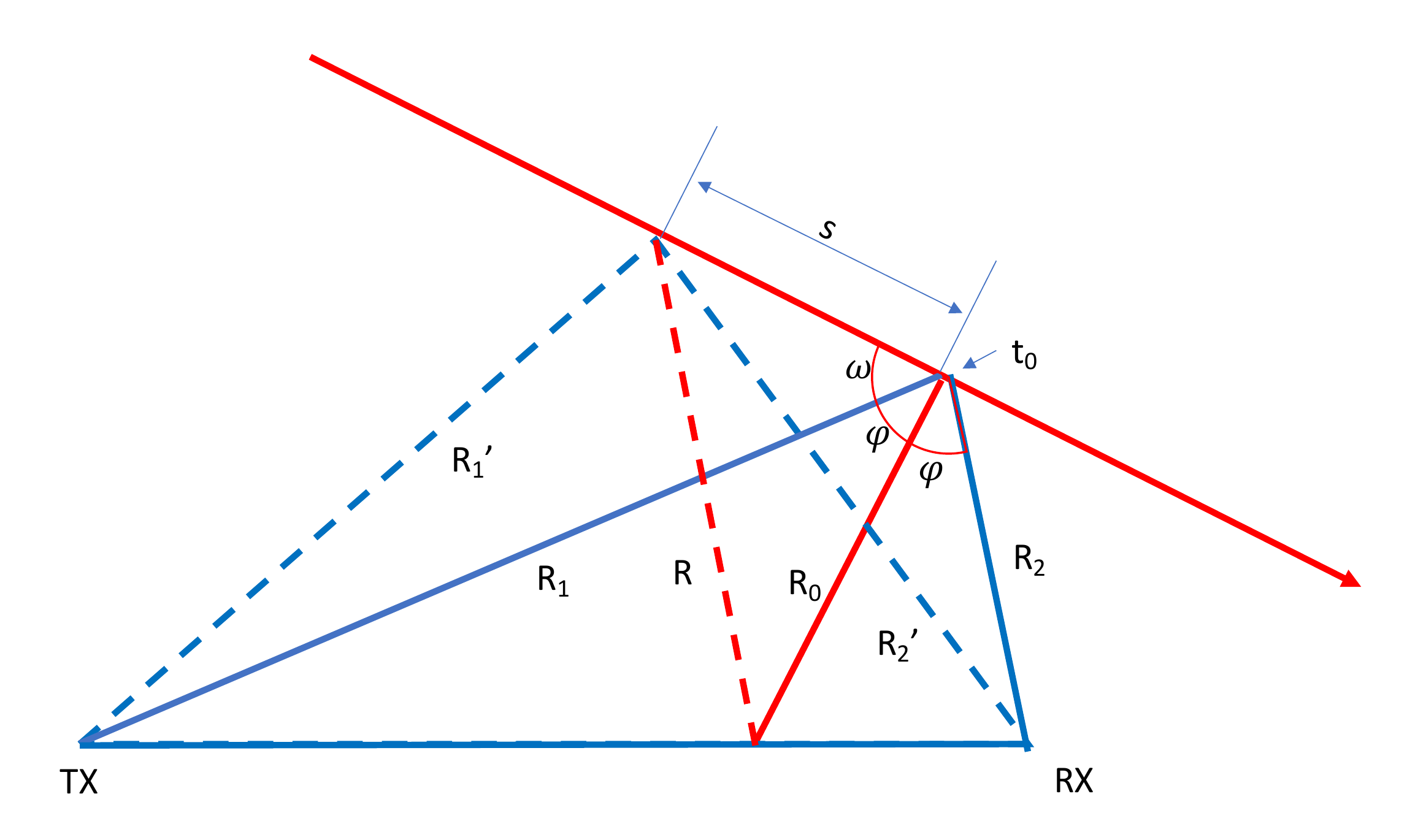}\centering
	\caption{[color on-line] TX and RX are the locations of the transmitter and receiver, respectively. The signal path for back scattering (red) and for forward scattering (blue) is shown for the optimal path corresponding to the reflection (specular) point, $t_0$, (full lines) and the path of a contributing neighboring point (dashed lines) separated by a distance $s$.  }\label{fig1}
\end{figure}

\begin{figure}
	\includegraphics[width=0.8\linewidth]{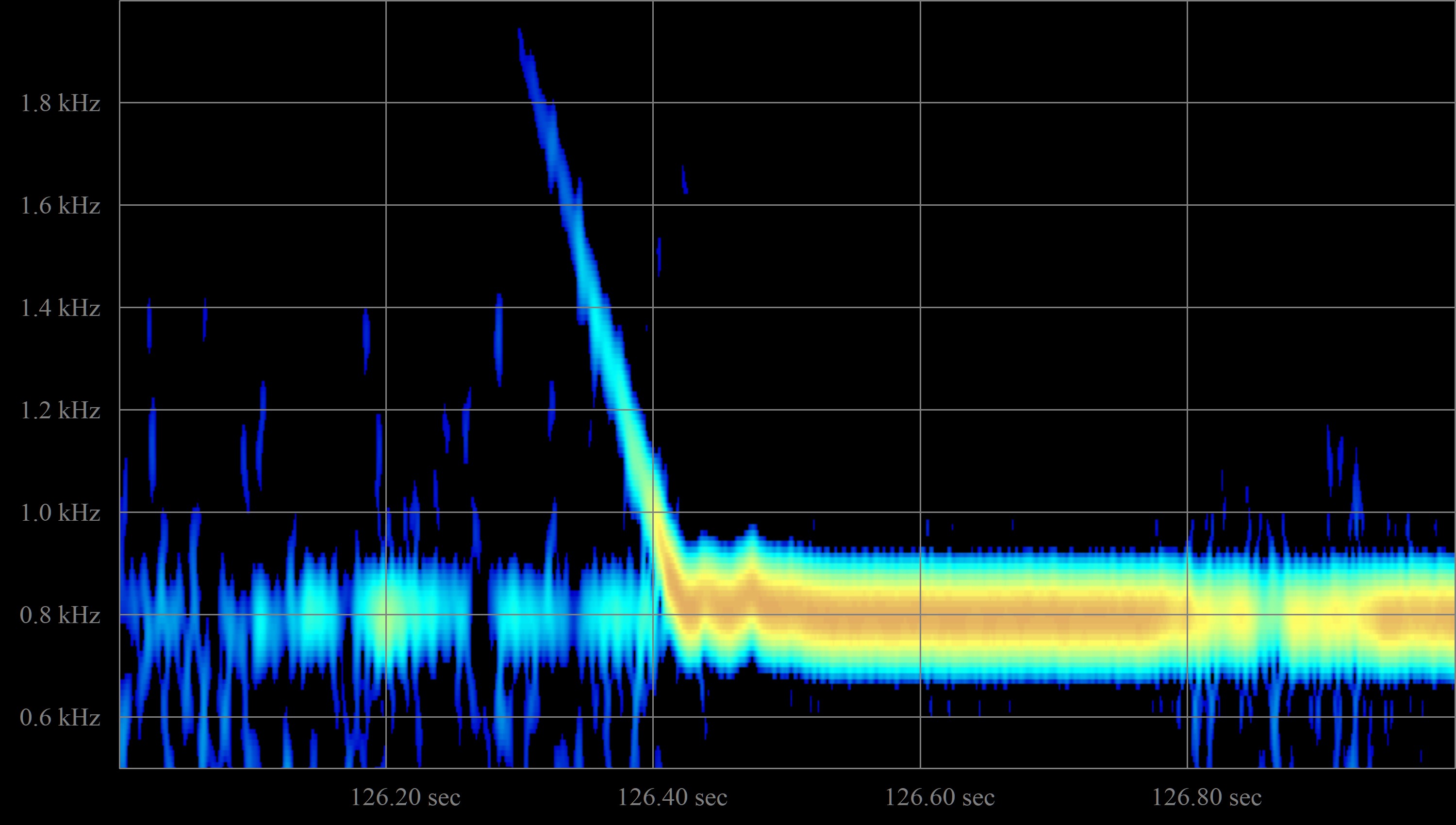}\centering
	\caption{Typical example, of a strong head echo. The horizontal scale is one second long, the vertical scale 1.5 kHz. The slope corresponds to -9.7 kHz/s Data were taken 05 May 2020 in Kampenhout (BE) using the 49990 kHz VVS beacon near Ieper (BE)\cite{Felix}}\label{ex1}.
\end{figure}

When an ionized trail is made by a meteoroid (see fig.~\ref{fig1}), the created  free electrons act as individual scatterers, they re-emit the signal of a transmitter in all directions. These can be observed in a receiver contributing an amplitude $dA_R$
\begin{equation}\label{dAr}
dA_R \propto \sin\left( 2\pi f t - \frac{2\pi(R'_1+R'_2)}{\lambda} \right) \ ,
\end{equation} where $\lambda$ is the transmitters wavelength and $f$  its frequency. ()All other parameters in this work are defined in fig.~\ref{fig1}.)
Each scattering contribution has a different phase depending on the distance $(R'_1+R'_2)$ the signal travels. Going along the trail the addition becomes  coherent when \nopagebreak $~{d (R'_1+R'_2)/dt=0}$.  At this point the path  followed  is a reflection,  the specular condition. The length of this path is $R_1+R_2$ and is the shortest path between transmitter and receiver via a point on the meteoroid trajectory. This point will be referred to as the specular point.
In fig.~\ref{fig1} also shows signal paths corresponding to back scattering. This is the radar setup, where emitter and transmitter are at nearly the same location. An arbitrary path has length $2R$ and the shortest path is $2R_0$. 
 Near the specular point where $s\ll R_0$  one finds that
\begin{eqnarray}
R_1'+R_2'&\approx &R_1 +R_2 +\frac{s^2 \sin^2 \omega}{2}\left (\frac{1}{R_1-s\cos\omega} +\frac{1}{R_2+s\cos\omega}\right )\\
&\approx& R_1 +R_2 +\frac{s^2 \sin^2 \omega}{2}\ \frac{R_1+R_2}{R_1 R_2} \ .\label{aberhallo}
\end{eqnarray}
This reduces to 
\begin{equation}\label{simple}
R \approx R_0 +\frac{s^2}{2R_0} 
\end{equation}
for back scattering. Also note that in chapter 9 of \cite{McK} the  notation was changed: $s \rightarrow f$ or \textit{\textsf{f}}. Here we will use $s$ consistently for the path of the meteoroid. Further note that, in general, $\omega \ne \pi/2-\phi$ because the trail may make an angle $\beta$ with the plane where forward scattering takes places, in which case  $ \sin^2\omega=1- \sin^2\phi\cos^2\beta$.

Central to the problem is the summation over individual scatterers along the trail. The  model assumes a constant ionization density along the trail, the amplitude is then given by
\begin{equation}\label{Integral}
A_R\propto\int_{x_1}^{x}\sin(\chi - \frac{\pi x^2}{2} )dx \ .
\end{equation}
Following \cite{McK}, we first evaluate the integral for back scattering  where $2s=x(R_0 \lambda)^{1/2}$ and $\chi=2\pi f t\  + $ a time independent phase. Thus one sums from the beginning of the trail until a point $s(t)$, i.e. the length of the trail at time t. Without loss of generality $x_1 \rightarrow -\infty$ can be assumed, as shown explicitly in section 3. One obtains
\begin{equation}
A_R \propto \left(C(x)+\frac{1}{2}\right) \sin \chi -\left(S(x)+\frac{1}{2}\right) \cos \chi \ ,
\end{equation}
where  $C$ and $S$ denote Fresnel integrals\footnote{Here $S(z)=\int_{0}^{z} \sin\frac{\pi t^2}{2} dt$ and $C(z)=\int_{0}^{z} \cos\frac{\pi t^2}{2} dt$}. To see what this means for the frequency we look at times $t<t_0$  and $t>t_0$  avoiding the complex behavior near $t_0$ by considering $|x|\gtrapprox1$.  With this approximation
\begin{equation}\label{aleft}
A_{<} \propto \frac{\cos (\chi- \frac{\pi  x^2}{2})}{\pi  x}\      \mathrm{and} \   A_{>}\propto\sin (\chi )-\cos (\chi )+\frac{\cos (\chi -\frac{\pi  x^2}{2})}{\pi  x}\  .
\end{equation}
  As $t$ approaches  $t_0$ the amplitude $A_<$ increases slowly compared with the frequency $f$. At each $t$ the instantaneous frequency $f_i$ can be obtained by determining  the phase $\Phi$ of $A_<(t)$
\begin{equation}\label{aleft2}
\Phi=\chi - \frac{\pi  x^2}{2} =2\pi f t - 2\pi \frac{s^2}{R_0\lambda}\ 
\end{equation}
and taking its derivative with respect to $t$, giving
\begin{equation}\label{fi}
f_i=\frac{1}{2\pi}\frac{d \Phi}{dt}=f-\frac{2s\ ds/dt}{R_0 \lambda} \ .
\end{equation}
Note that here $s>0$ and $ds/dt <0$, the shift is thus positive. In practice one analyses the change in the instantaneous frequency $f_i$,
\begin{equation}\label{timedepshift}
\frac{df_i}{dt}=-2\left( \left[ \frac{ds}{dt} \right]^2 + s\frac{d^2s}{dt^2} \right)\frac{1}{R_0 \lambda} \ .
\end{equation}
A simple model choice is a constant velocity where  $s=|V (t_0- t)|$, so that 
\begin{equation}\label{8-15r}
\frac{df_i}{dt}=-\frac{2V^2}{R_0\lambda}\  .
\end{equation}
To obtain the shift in forward scattering one simply replaces eq.~\ref{simple} wit eq.~\ref{aberhallo} to find
\begin{equation}\label{result}
\frac{df_i}{dt}\approx-\frac{V^2 \sin^2\omega}{\lambda}\frac{R_1+R_2}{R_1R_2}\ .
\end{equation}

For $t>t_0$ there is no such simple derivation for the phase of $A_>$ possible. In order to get insight numerical calculations were done. These  will be discussed in section~\ref{calcs}.

\section{Derivation of the frequency shift in terms of Doppler shifts}
\begin{figure}\centering
	\includegraphics[width=0.8\linewidth]{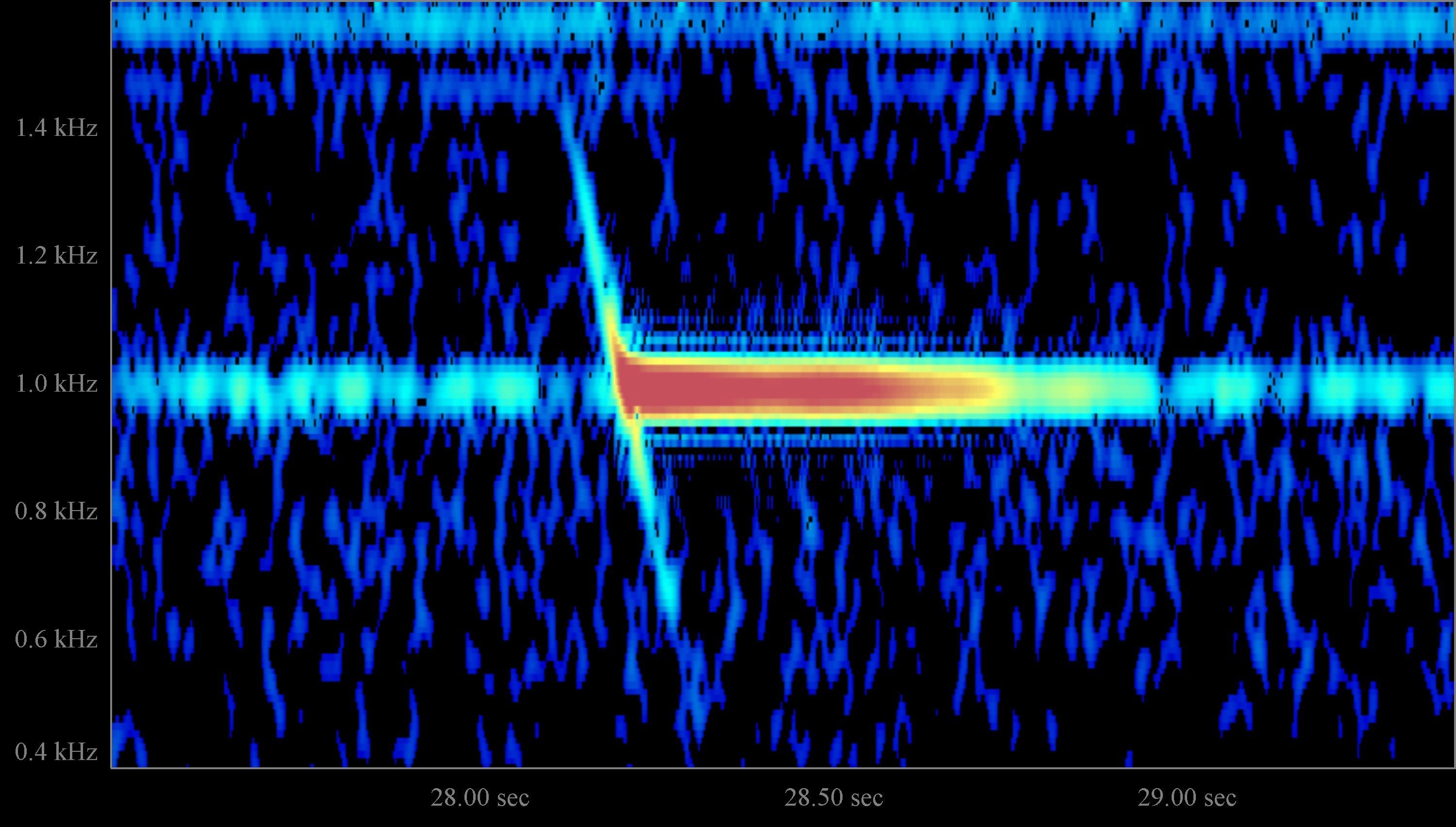}
	\caption{Less common head echo showing a shift that extends to negative shifts. The slope is -4.9 kHz/s. Data taken on  09 June 2011 in Kampenhout using the  49970 kHz BRAMS beacon at Dourbes (BE)\cite{Felix}}\label{ex2}.
\end{figure}

In this section we consider the possibility that only a short part of the trail survives, so that it has a length $x <1$. Thus where at the front free electrons are created they disappear at the back. Therefore, the trail has a constant length $\Delta x$. In fact for the observer it appears as a passing object, which maybe as well be the meteoroid itself. When this object passes point $t_0$ it will give an echo. In this case one can approximate eq.~\ref{Integral} by
\begin{equation}\label{delta}
A_R\propto\sin(\chi - \frac{\pi x^2}{2} )\Delta x \ .
\end{equation}
The phase of this amplitude is identical to that in eq.~\ref{aleft2} and the same relations hold for the shifts. 
 The shift continues for $t>0$ where the instantaneous  frequency, $f_i$, is lower than the emitter frequency $f$.  
 
 The bistatic Doppler shift is given by $-\frac{1}{\lambda}{d (R'_1+R'_2)/dt}$. Using the same approximations as eq.~\ref{aberhallo}ione arrives at the identical expression  as in eq.~\ref{fi}. Thus the shifts are the same as they must be, because of Galilean invariance.  A more informative evaluation will come from a numerical calculation in the next section.

\section{Example Calculations}\label{calcs}
At this point it may be interesting to consider in more detail how the signal appears in observation. The result of section 2 and 3 can be generalized into one expression
\begin{equation}\label{ARl}
A_R \propto \left(C(x)-C(x-\Delta x)\right) \sin \chi -\left(S(x)-S(x-\Delta x )\right) \cos \chi \ ,
\end{equation}
where $\Delta x$ is the length of the object or trail as defined above. In the following calculation we use for $s$ the backscatter configuration with $V=30$ km/s, $R_0=100$ km, and an emitter frequency $f=50$~MHz.

\begin{figure}\centering
	\includegraphics[width=0.7\linewidth]{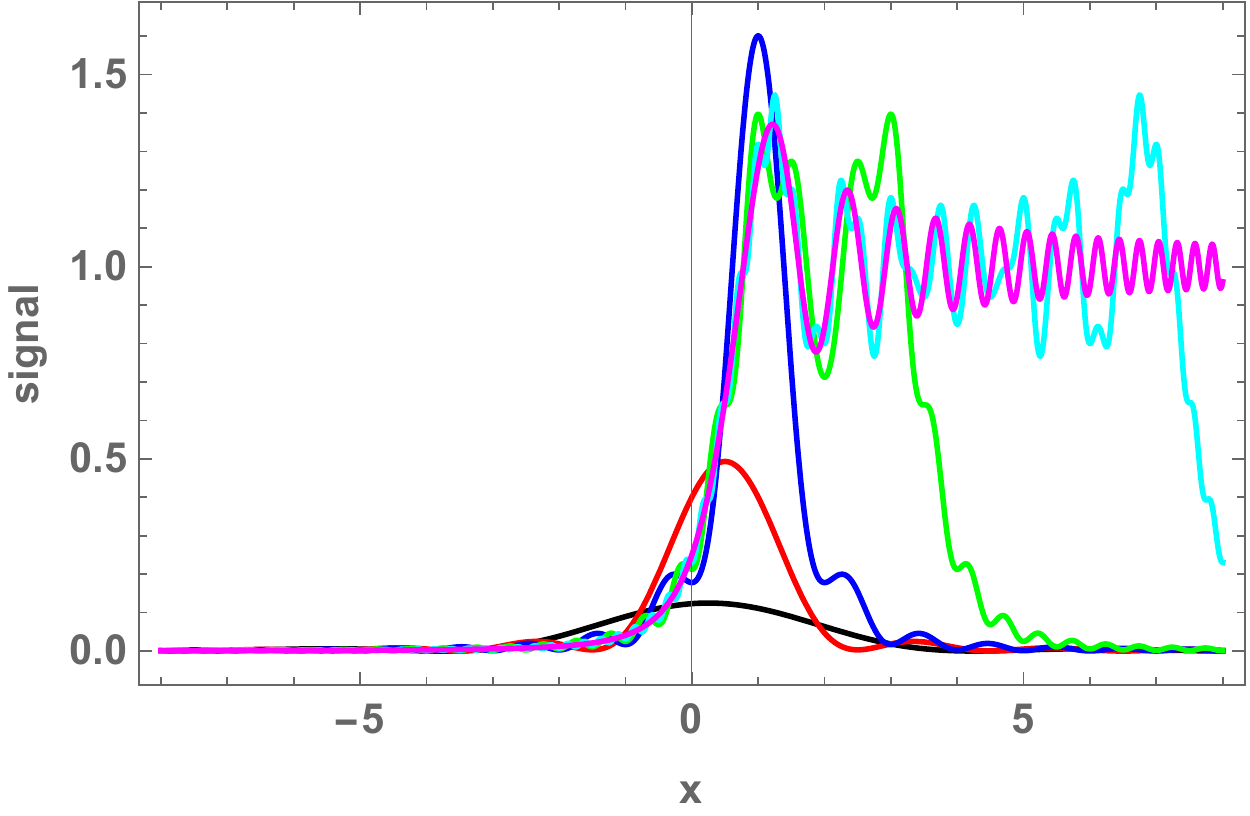}
	\caption{$\overline{  A_R^2(x(t))}$ for $\Delta x= 0.5,1,2,4,8,\infty$ (black, red, blue, green, cyan, magenta)}\label{powerplot}
\end{figure}

First we consider the signal power,
$\overline{  A_R^2(t)}$, which is obtained by averaging over a time long with respect to the frequency but small with respect to $x(t)$. In fig.~\ref{powerplot} this quantity is shown, evaluated for
various values of $\Delta x$.
 For small $\Delta x$ the signal has a nearly Gaussian dependence, where it should be noted that $x=1$ corresponds to about one Fresnel zone (here it   corresponds to $t=\sqrt{\lambda R_0}/2V=13\ \mathrm{ms}$ or a distance of $387\  \mathrm{m}$). At this small distance  the signal has the characteristics of Fraunhofer slit scattering.  For $\Delta x>2 $  and $x>1$ one  observes what are called Fresnel oscillations. $\Delta x=\infty$ refers to the situation discussed in section 2; the pattern seen here  corresponds to Fresnel edge scattering. The oscillation frequency in the Fresnel pattern (see chapter 8 eq.~(8-15) in \cite{McK}) is identical to the frequency shift calculated in section 2. In fact, historically, it has been a main tool for determining meteoroid velocities instead of the frequency shift.
 \begin{figure}
 	\includegraphics[width=0.4\linewidth]{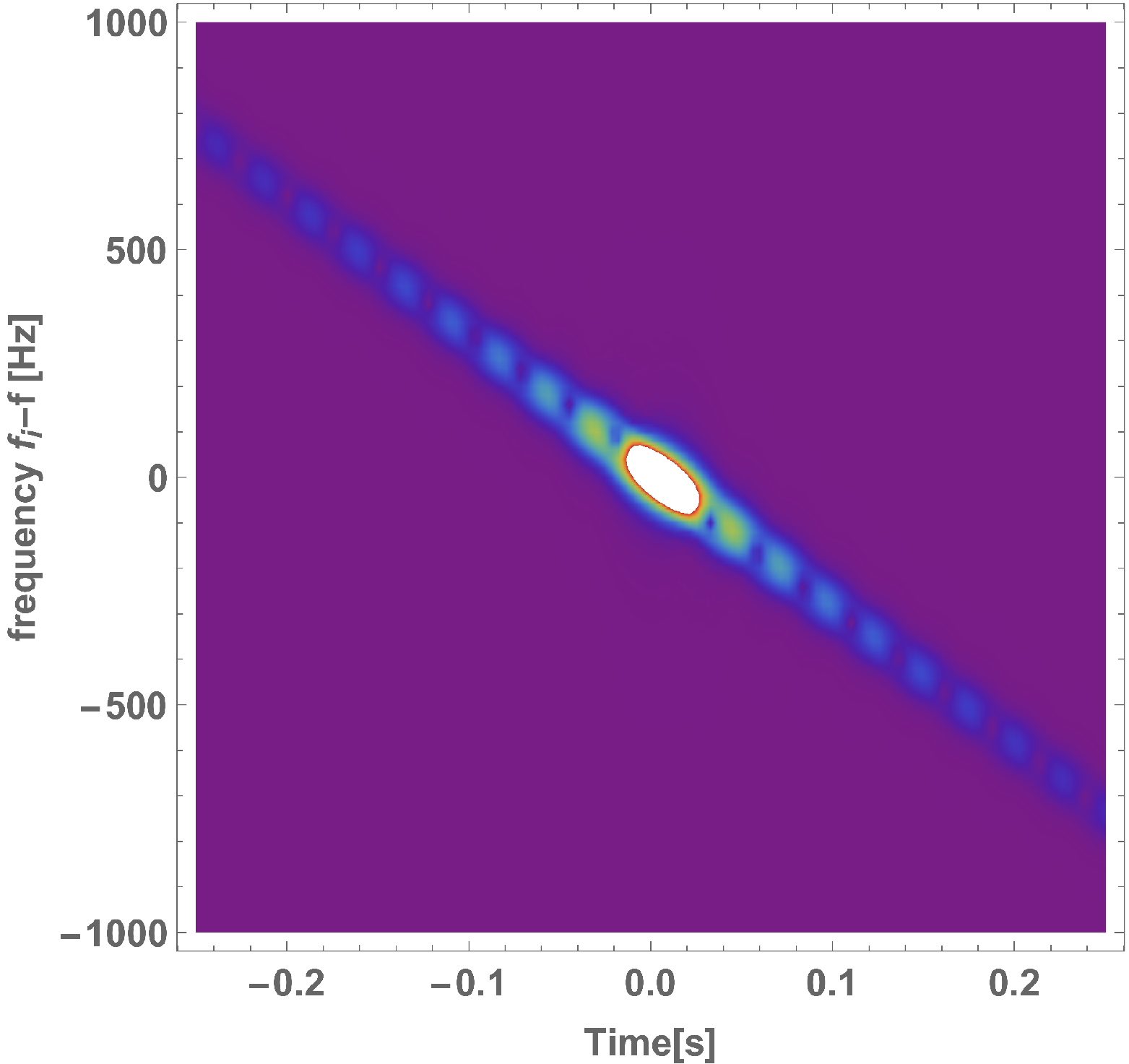}\hfill	\includegraphics[width=0.4\linewidth]{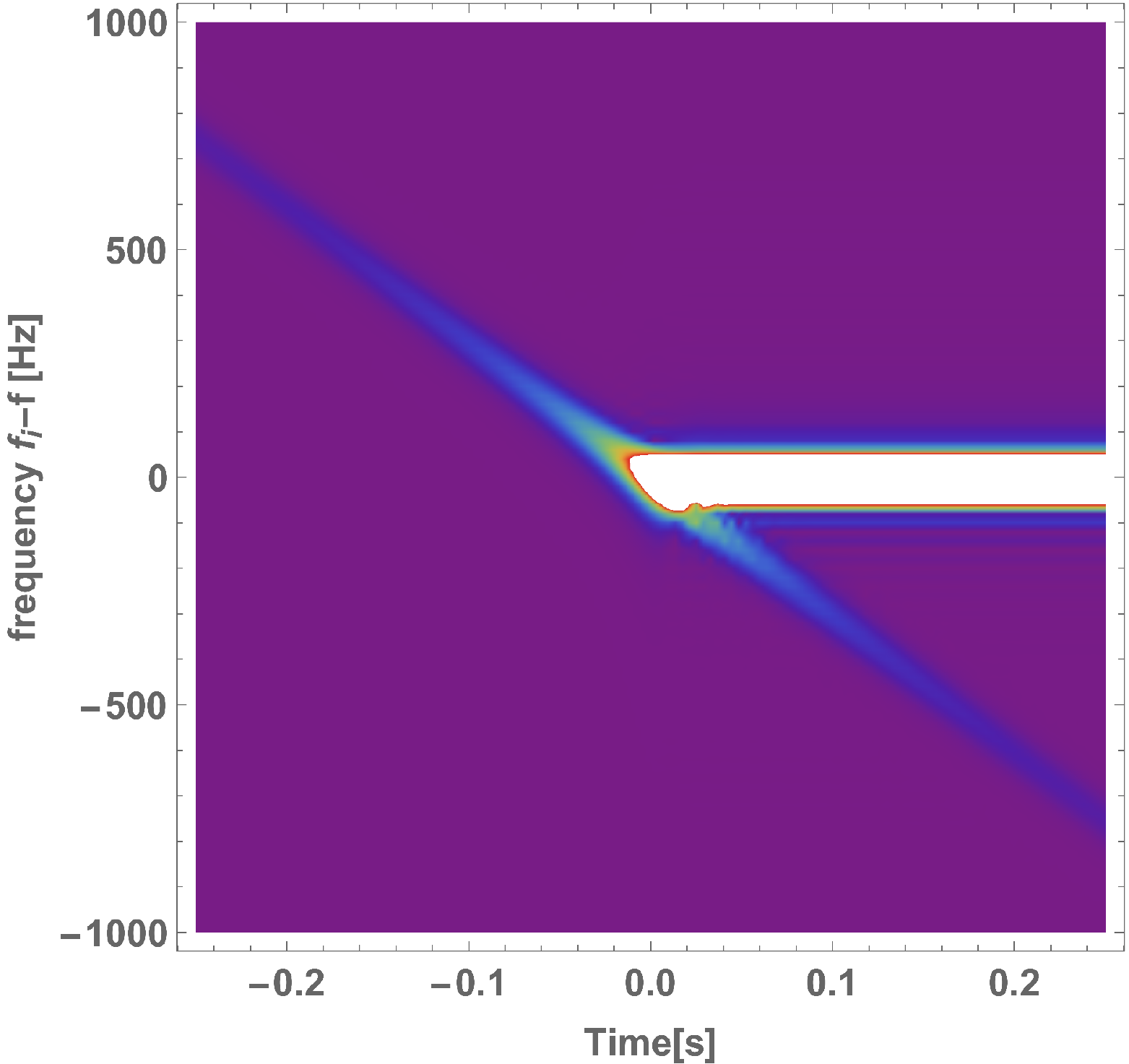}
 	\caption{[Color online] Time versus frequency of $\log(A_R)$ obtained with a STFT procedure. Left: passing trail or object. Right: Dynamic meteoroid trail. See text}\label{spectra}
 \end{figure}

 To observe the frequency dependence of eq.~\ref{ARl}, it should be displayed as a time dependent frequency spectrum similar to the observed data. This can be done by a Short Time Fourier Transform (STFT)  of the theoretical expressions for $A_R$ in eq.~\ref{ARl} (see appendix \ref{appa}). The results are shown in fig.~\ref{spectra}. First we discuss the spectrum (left) which is close to what was discussed in the previous section. The size of the object or trail passing through the specular point is taken to be $\Delta x=1$.  The signal has the characteristic Doppler shift dependence with a maximum at the specular point. Superimposed is a Fraunhofer diffraction spectrum. This pattern has to occur independent of the interpretation of the object in terms of a a short-lived trail or as moving object. The spectrum on the right is for $\Delta x=\infty$, the line oscillator model of section 2. It is dominated by the ridge  at $f_i - f=0$ at $t>t_0$. Its origin is the stationary  trail. The diagonal in the spectrum has the same time-frequency dependence as  the  left-hand spectrum but without the diffraction pattern. This is the head echo from the trail as it is being formed. 
 
\begin{figure}
 	\includegraphics[width=0.8\linewidth]{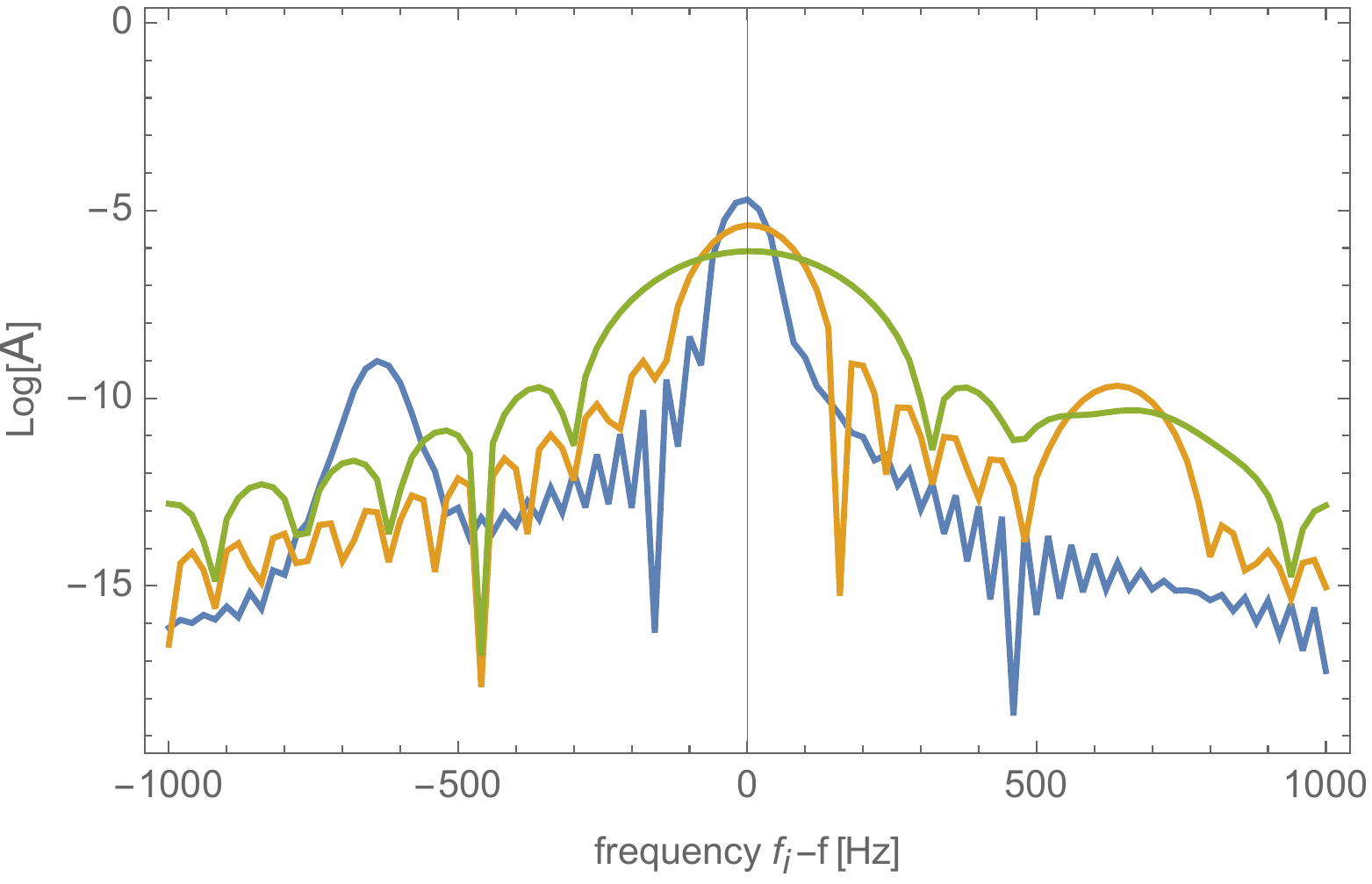}\	
 	\caption{[Color online]  Frequency spectrum of $\log(\mathcal{A_R})$ at $t=0.2$~s the green, orange and blue curves refer to an average over 0.5, 1, and 2 Fresnel zones, respectively. The branch of the head echo is at -700 Hz. See also text}\label{three}
 \end{figure}

 There are several  observations to be made at this point: The Fresnel oscillations found in fig.~\ref{powerplot} are not seen in the spectrum. They  appear by reducing  the Hann window in the STFT (see appendix \ref{appa}). Calculations were made for a window width of 0.5, 1 and 2 Fresnel zones. For the short window the oscillations can be seen, but at the cost of resolution in frequency. This is characteristic for a Fourier transform where  time and frequency resolution exclude each other mutually.  In fig.~\ref{three} the frequency spectrum is shown at $t=0.2$~s for the three Hann windows. The spectrum  with the largest  Hann window was also used to obtain fig.~\ref{spectra}. The branch of the head echo is clearly seen for that window, while for the smallest window it has all but disappeared due to the reduced frequency resolution.  Notice that the head echo signal is more than 100 times smaller compared to the signal of the stationary trail. This indicates that measuring the head echo at $t>t_0$   requires that the receiver settings  and the associated analysis programs  need to take the time average in consideration. The observation of the frequency shift at $t<t_0$ is rather robust and thus also in an actual measurement. If possible, one could determine both the Fresnel oscillations and the frequency shift by analyzing the data in two different ways by good time and poor frequency resolution and vice versa, respectively. In this way one has  two ways to measure the meteoroid velocity.

 One might expect measured events as in fig.~\ref{ex2} to be the most common. However, most observations are as in fig.~\ref{ex1}. This is either due to problems associated with measuring the head exho at $t>t_0$ for the reasons mentioned above or because the line-scattering model is not describing the meteoroid events adequately.  Another reason for this can be the observational bias for strong echo's with a short trail below the specular point. 

 \section{The half-ball model}
\begin{figure}\centering 
	\includegraphics[width=0.7\linewidth]{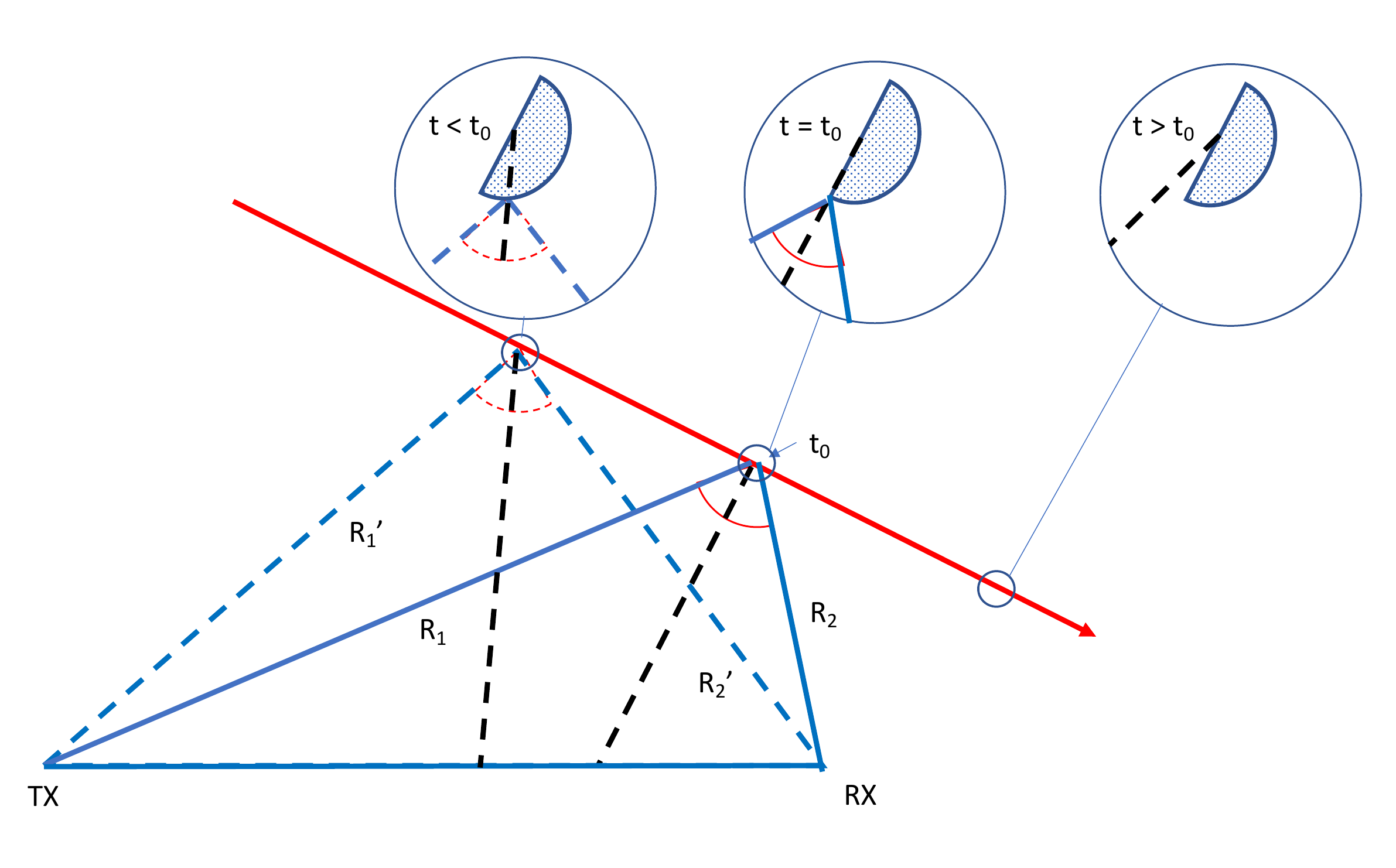}\hfill	\includegraphics[width=0.3\linewidth]{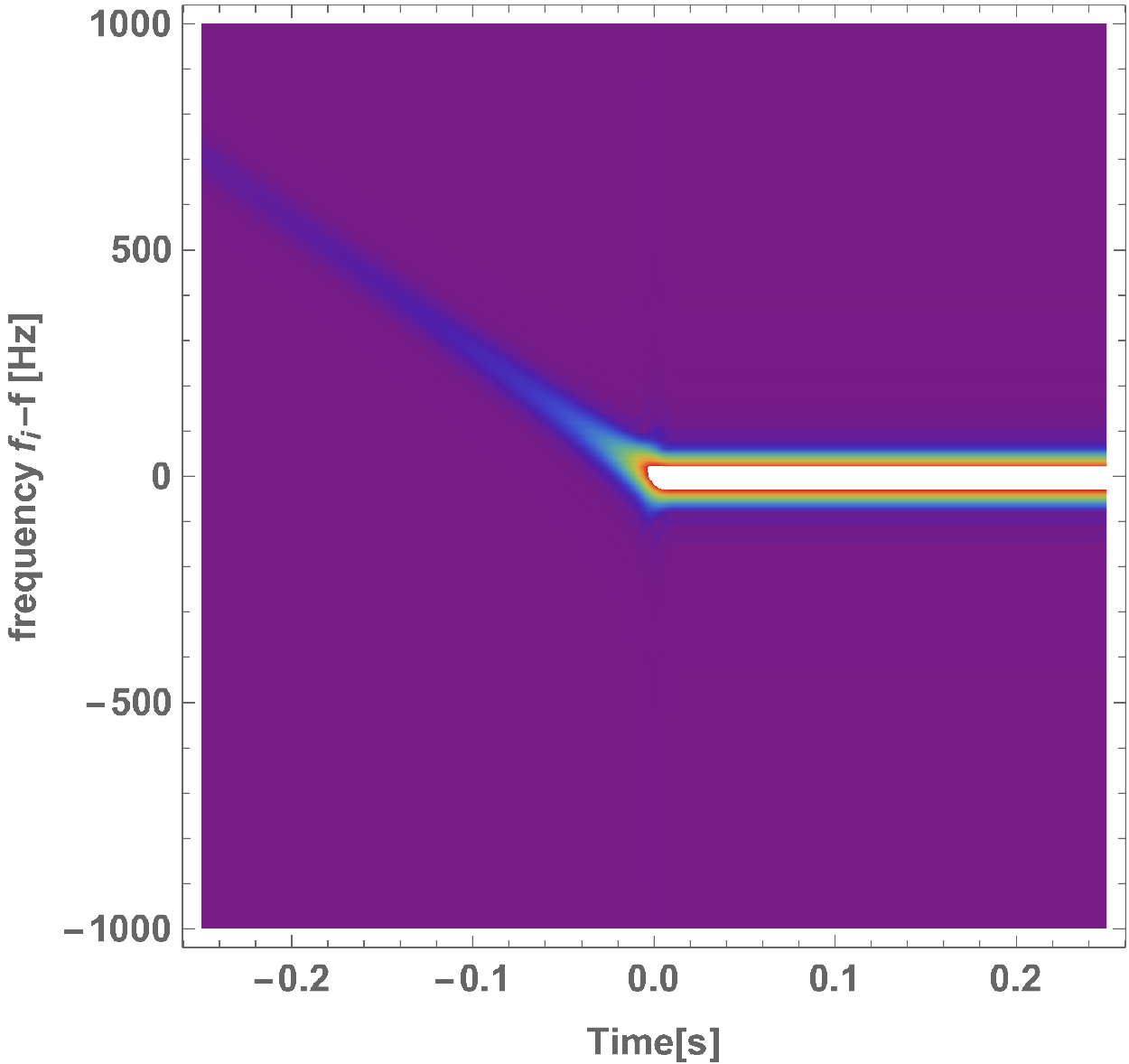}
	\caption{[Color online] Half-ball model. An echo can be seen at any point along the meteoroid trajectory. But only 
		 from the front plasma of the meteoroid. After passing the specular point ($t=t_0$), there will be no echo any longer except from the stationary trail. The right hand side shows a calculation where this was implemented in the line-oscillator model although this is not the right model for such a scenario. }\label{halfcanon}
\end{figure}

The characteristic asymmetric behavior of the head echo noticed at the end of the previous section can be resolved assuming  reflections from the   plasma in front of -- and moving with -- the meteoroid.  Research around 2000, for example with Arecibo \cite{Mathews} and ALTAIR \cite{NASA} shows the importance of the head echo plasma. Such a plasma reflects radio waves  like a metal mirror. The specular condition is then irrelevant since there will always be a spot on a sphere allowing reflection from transmitter to receiver; also in forward scattering. But in forward scattering it is important to realize that  only the front half of the meteoroid has this property. The situation is sketched in fig.~\ref{halfcanon}. Only in the approaching phase reflections are possible up to the specular point, after that reflections would have to come from the back of the meteoroid. Therefore, in this scenario a true Doppler shift signal is observed until $t=t_0$, and after that only the trail left by the meteoroid at the  specular point contributes to the signal. This scenario, which could  be called the "half-ball model", would explain experimental data as in fig.~\ref{ex1}.  Reflections at the specular point proceed via the shortest path between transmitter and receiver and therefore  also give the strongest reflection in this model, as it does in the line-oscillator model. It will be interesting to combine line-oscillator model (observing the back tail) with the "half-ball model" as the relevant trail parts do not pass the specular point at exactly the same time. In any case it  is somewhat surprising that the head plasma  reflection is strong enough to be seen in a comparatively modest amateur setup. 

\section{Conclusions}
Two very different approaches lead to the same shifts in the radio-echo frequency when a meteoroid passes through a point where the specular condition is fulfilled. One assumes either thermalized electrons in the local atmosphere \label{key}or electrons co-moving with the meteoroid as the scatterers. They lead to the same frequency shifts because they are both based on the time derivative of eq.~\ref{aberhallo}. 
The model discussed in section 2 describes head echos and the resulting stationary trail in a unified way. More explicitly, further calculations in section~\ref{calcs} show that frequency shift and the Fresnel diffraction pattern relates to the same observable eq.~\ref{fi}.
In section 3 it was shown that a short-lived trail and a moving object give the same frequency shift  of the head echo  as in section 2. The size of the object  must have $|x|\lessapprox1$, i.e one Fresnel zone. 
The calculations show that in these cases a diffraction pattern should be visible on the head echo. In practice this is not observed. This may be because of the simplicity of the model.
 The calculations in section~\ref{calcs} allow for a qualitative comparison with actual observations. The contradicting requirements for frequency and time resolution were pointed out and it was also shown that this is  reflected in the duality observing either the Fresnel diffraction or the frequency shifts. This duality  may also play a role in the absence of a head echo  at $t>0$ in most observations. Another reason for this can be the observational bias for strong echo's with a short trail below the specular point. However, assuming that a co-moving plasma in front of the meteoroid is important, there is a simple geometric argument why there is no head echo after the meteoroid passes the specular point $t_0$.


As a final remark and recommendation: The measurements of meteoroid echos contain much more information than the frequency shifts. In particular the rapid decay of the signal after the head echo, occurring within a second, may contain more hints as to the precise nature of the meteoroid event. Modern equipment allows a quantitative measurement of the signal strength and the classic text of \cite{McK} provides several methods for its analysis.
\appendix 
\section{Appendix STFT}\label{appa}The Short Time Fourier Transform has here the following form
\begin{eqnarray*}\label{STFT}
\mathcal{A}_R(f_i-f,t)&=&\int_{-\infty}^{\infty}A_R((2\pi f  (t+t'), x(t+t'))e^{i2\pi(f_i-f)(t+t')}\mathrm{HannWindow}(\alpha t')dt'\ ,\\
\mathrm{HannWindow}(w)&=& 
\begin{array}{cc}
\left\{
\begin{array}{cc}
\frac{1}{2}+\frac{1}{2} \cos (2 \pi w)& -\frac{1}{2}\leq w\leq \frac{1}{2} \\
 & \\
0 & | w| >\frac{1}{2} \\
\end{array}\right .
\end{array} \ ,
\end{eqnarray*}
where $\alpha$ determines the width of the window.
\section*{Acknowledgement}
The author thanks F. Verbelen, W. Kaufmann, and M.T. German for clarifying some of the concepts in radio meteoroid measurements, for sharing their data, and for discussions.

\end{document}